\documentclass[10pt, conference]{IEEEtran}
\usepackage{fancyhdr}

\usepackage[T1]{fontenc}
\usepackage{color}
\usepackage{graphicx}
\usepackage{subcaption}
\usepackage{amsmath}
\usepackage{amssymb}
\usepackage[colorlinks=true,urlcolor=blue,citecolor=blue,linkcolor=blue,bookmarks=true]{hyperref}
\usepackage{multirow}
\usepackage{multicol}

\newcommand{\singleprecefficiency}{90.7}
\newcommand{\singleprecperfsust}{325.8}

\newcommand{\halfprecefficiency}{90.7}
\newcommand{\halfprecperfsust}{999.0}
\newcommand{\halfprecperfsustef}{1.0}

\newcommand{\halfprecperfpeakef}{1.13}

\newcommand{\heronodecount}{4560}
\newcommand{\herogpucount}{27360}
\IEEEoverridecommandlockouts
\begin{document}

% make the title area

\title{Exascale Deep Learning for Climate Analytics}
%\title{Towards Exascale Deep Learning:\\ Analysis of Extreme Weather Patterns at 263 PF/s}

\newif\ifprintauthors

%Comment this line to suppress printing authors
\printauthorstrue

\ifprintauthors
 \author{
 \IEEEauthorblockN{Thorsten Kurth\IEEEauthorrefmark{1}}
 \IEEEauthorblockA{%NERSC\\
 %Lawrence Berkeley National Laboratory\\
 %Berkeley, CA 94720, USA\\
 tkurth@lbl.gov}\\
 \IEEEauthorblockN{Mayur Mudigonda\IEEEauthorrefmark{1}}
 \IEEEauthorblockA{%NERSC\\
 %Lawrence Berkeley National Laboratory\\
 %Berkeley, CA 94720, USA\\
 mudigonda@berkeley.edu}\\
  \IEEEauthorblockN{Ankur Mahesh\IEEEauthorrefmark{1}}
 \IEEEauthorblockA{%NERSC\\
 %Lawrence Berkeley National Laboratory\\
 %Berkeley, CA, 94720, USA\\
 amahesh@lbl.gov}\\
 \IEEEauthorblockN{Massimiliano Fatica\IEEEauthorrefmark{2}}
 \IEEEauthorblockA{%NVIDIA\\
 %Santa Clara, CA 95051, USA\\
 mfatica@nvidia.com}\\
 \IEEEauthorblockA{
 \IEEEauthorrefmark{1}Lawrence Berkeley National Laboratory \\
 Berkeley, CA 94720, USA
 }
 \and
 \IEEEauthorblockN{Sean Treichler\IEEEauthorrefmark{2}}
 \IEEEauthorblockA{%NVIDIA\\
 %Santa Clara, CA 95051, USA\\
 sean@nvidia.com}\\
 \IEEEauthorblockN{Nathan Luehr\IEEEauthorrefmark{2}}
 \IEEEauthorblockA{%NVIDIA\\
 %Santa Clara, CA 95051, USA\\
 nluehr@nvidia.com}\\
 \IEEEauthorblockN{Michael Matheson\IEEEauthorrefmark{3}}
 \IEEEauthorblockA{%OLCF\\
 %Oak Ridge National Laboratory\\
 %Oak Ridge, TN 37831, USA\\
 mathesonma@ornl.gov}\\
 \IEEEauthorblockN{Prabhat\IEEEauthorrefmark{1}}
 \IEEEauthorblockA{%NERSC\\
 %Lawrence Berkeley National Laboratory\\
 %Berkeley, CA 94720, USA\\
 prabhat@lbl.gov} \\
 \IEEEauthorblockA{
 \IEEEauthorrefmark{2}NVIDIA\\
 Santa Clara, CA 95051, USA
 }
 \and
 \IEEEauthorblockN{Joshua Romero\IEEEauthorrefmark{2}}
 \IEEEauthorblockA{%NVIDIA\\
 %Santa Clara, CA 95051, USA\\
 joshr@nvidia.com} \\
 \IEEEauthorblockN{Everett Phillips\IEEEauthorrefmark{2}}
 \IEEEauthorblockA{%NVIDIA\\
 %Santa Clara, CA 95051, USA\\
 ephillips@nvidia.com}\\
 \IEEEauthorblockN{Jack Deslippe\IEEEauthorrefmark{1}}
 \IEEEauthorblockA{%NERSC\\
 %Lawrence Berkeley National Laboratory\\
 %Berkeley, CA 94720, USA\\
 jrdeslippe@lbl.gov}\\
 \IEEEauthorblockN{Michael Houston\IEEEauthorrefmark{2}}
 \IEEEauthorblockA{%NVIDIA\\
 %Santa Clara, CA 95051, USA\\
 mhouston@nvidia.com} \\
 \IEEEauthorblockA{%OLCF\\
 \IEEEauthorrefmark{3}Oak Ridge National Laboratory\\
 Oak Ridge, TN 37831, USA
 }
 }
 \author{
 \IEEEauthorblockN{Thorsten Kurth\IEEEauthorrefmark{1}}
 \IEEEauthorblockA{%NERSC\\
 \thanks{\IEEEauthorrefmark{1}
 Lawrence Berkeley National Laboratory, 
 Berkeley, CA 94720, USA }
 tkurth@lbl.gov}\\
 \IEEEauthorblockN{Nathan Luehr\IEEEauthorrefmark{2}}
 \IEEEauthorblockA{%NVIDIA\\
 %Santa Clara, CA 95051, USA\\
 nluehr@nvidia.com}\\
 \IEEEauthorblockN{Jack Deslippe\IEEEauthorrefmark{1}}
 \IEEEauthorblockA{%NERSC\\
 %Lawrence Berkeley National Laboratory\\
 %Berkeley, CA 94720, USA\\
 jrdeslippe@lbl.gov}\\
 \and 
 \IEEEauthorblockN{Sean Treichler\IEEEauthorrefmark{2}}
 \IEEEauthorblockA{%\\
 \thanks{\IEEEauthorrefmark{2}
 NVIDIA,
 Santa Clara, CA 95051, USA }
 sean@nvidia.com}\\
 \IEEEauthorblockN{Everett Phillips\IEEEauthorrefmark{2}}
 \IEEEauthorblockA{%NVIDIA\\
 %Santa Clara, CA 95051, USA\\
 ephillips@nvidia.com}\\
 \IEEEauthorblockN{Massimiliano Fatica\IEEEauthorrefmark{2}}
 \IEEEauthorblockA{%NVIDIA\\
 %Santa Clara, CA 95051, USA\\
 mfatica@nvidia.com}\\
 \and
 \IEEEauthorblockN{Joshua Romero\IEEEauthorrefmark{2}}
 \IEEEauthorblockA{%NVIDIA\\
 %Santa Clara, CA 95051, USA\\
 joshr@nvidia.com} \\
 \IEEEauthorblockN{Ankur Mahesh\IEEEauthorrefmark{1}}
 \IEEEauthorblockA{%NERSC\\
 %Lawrence Berkeley National Laboratory\\
 %Berkeley, CA, 94720, USA\\
 amahesh@lbl.gov}\\
 \IEEEauthorblockN{Prabhat\IEEEauthorrefmark{1}}
 \IEEEauthorblockA{%NERSC\\
 %Lawrence Berkeley National Laboratory\\
 %Berkeley, CA 94720, USA\\
 prabhat@lbl.gov} \\
 \and
 \IEEEauthorblockN{Mayur Mudigonda\IEEEauthorrefmark{1}}
 \IEEEauthorblockA{%NERSC\\
 %Lawrence Berkeley National Laboratory\\
 %Berkeley, CA 94720, USA\\
 mudigonda@berkeley.edu}\\
 \IEEEauthorblockN{Michael Matheson\IEEEauthorrefmark{3}}
 \IEEEauthorblockA{%\\
 \thanks{\IEEEauthorrefmark{3}
 Oak Ridge National Laboratory,
 Oak Ridge, TN 37831, USA }
 mathesonma@ornl.gov}\\
 \IEEEauthorblockN{Michael Houston\IEEEauthorrefmark{2}}
 \IEEEauthorblockA{%NVIDIA\\
 %Santa Clara, CA 95051, USA\\
 mhouston@nvidia.com} \\
 }
 \IEEEaftertitletext{\vspace{-1cm}}
\fi

\maketitle
\thispagestyle{fancy}
\lhead{}
\rhead{}
\chead{}
\lfoot{\footnotesize{
SC18, November 11-16, 2018, Dallas, Texas, USA
\newline 978-1-5386-8384-2/18/\$31.00 \copyright 2018 IEEE}}
\rfoot{}
\cfoot{}
\renewcommand{\headrulewidth}{0pt}
\renewcommand{\footrulewidth}{0pt}

% add page numbers for the draft - remove for a camera-ready submission
%\pagestyle{plain}
%\thispagestyle{plain}

%\vspace*{0.25cm}
\begin{abstract}
%\section{Abstract}

%\textcolor{red}{150 words max}
%We use Deep Learning on the two largest GPU supercomputers available today  to address a challenging problem in science. 
%We examine extraction of pixel-level masks of extreme weather patterns using a residual neural network based on Tiramisu, and implemented in TensorFlow and Horovod. We describe improvements to both the software frameworks, input pipeline,  and the network training algorithms necessary to efficiently scale deep learning to thousands of GPUs on the two largest GPU supercomputers available today. The Tiramisu network scales to 5300 P100 GPUs (the full Piz Daint system)  with  a sustained throughput of 18.6 PetaFLOP/s and parallel efficiency of 78.9\%  and to 15360 V100 GPUs on the Summit supercomputer with a sustained  throughput of 108.3 PetaFLOP/s and a parallel efficiency of 88.0\%. By taking advantage of the FP16 performance on the V100 GPUs, a half-precision version of the Tiramisu network achieves a peak and sustained throughput of 263.2 and 232.9 PetaFLOP/s respectively on 15360 GPUs. The network can train to convergence in 100 minutes.
%The present work and the proposed improvements to popular frameworks (TensorFlow and Horovod)  open new frontiers for applying deep learning at unprecedented scale for science.

We extract pixel-level masks of extreme weather patterns using variants of Tiramisu and DeepLabv3+ neural networks. We describe improvements to the software frameworks, input pipeline, and the network training algorithms necessary to efficiently scale deep learning on the Piz Daint and Summit systems. 
The Tiramisu network scales to 5300 P100 GPUs with a sustained throughput of 21.0 PF/s and parallel efficiency of 79.0\%. 
DeepLabv3+ scales up to \herogpucount\, V100 GPUs with a sustained  throughput of \singleprecperfsust\, PF/s and a parallel efficiency of \singleprecefficiency\% in single precision. By taking advantage of the FP16 Tensor Cores, a half-precision version of the DeepLabv3+ network achieves a peak and sustained throughput of \halfprecperfpeakef\, EF/s and \halfprecperfsust\, PF/s respectively. 
%Our optimizations are implemented in open-source TensorFlow and Horovod frameworks, and will benefit the broader community.

\end{abstract}

%\st{The network can train to convergence in 100 minutes.}

\vspace*{-0.1cm}

\section{Justification}
\label{sec:justification}
\vspace*{-0.25cm}

%\textcolor{red}{50 words max}
We apply segmentation architectures to climate datasets; achieving state-of-the-art weather pattern masks. We scale the architectures to \herogpucount\, Volta GPUs, obtaining a peak (sustained) FP16 performance of \halfprecperfpeakef\, EF/s (\halfprecperfsustef\, EF/s). We developed methodologies at system level and several deep learning algorithmic innovations to achieve this unprecedented scaling. 
\section{Performance Attributes}
\label{sec:attributes}

\begin{center}
  \begin{tabular}{p{4cm}p{4cm}}
    \hline
    Performance Attribute & Our submission \\ \hline \hline
    Category of Achievement & Peak performance, Time-to-solution\\
    Type of Method Used & Deep Learning \\
    Results reported on basis of & Whole application \\
                       & including I/O\\
    Precision reported & Mixed precision \\
    System scale & Measured on full system \\
    Measurement mechanism & Application timers \\ \hline
  \end{tabular}
\end{center}

%\vspace*{-1.5cm}

\section{Overview}
\label{sec:science-drivers}

%\textcolor{red}{description of the problem and its importance, in terms understandable to a non-specialist (1 p max)}

\subsection{Pattern Detection for Characterizing Extreme Weather}

Climate change poses a major challenge to humanity in the 21st century. Several nations are considering adaptation and mitigation strategies pertaining to global, mean quantities such as temperature, or sea-level rise. Increasingly, state and local governments are interested in the question of how extreme weather events will change and affect their local communities. For instance, the state of California receives over 50\% of its rainfall through Atmospheric Rivers (ARs), and Water Resource Management planners are interested in understanding if AR tracks will shift in the future, potentially resulting in a dramatic shortfall in fresh water supply. In the state of Florida, homeowners are interested in understanding if Tropical Cyclones (TCs) or hurricanes will become more intense and start making landfall more often. This has a direct impact on home prices and the insurance industry. TCs have caused the US economy over \$200B worth of damage in 2017, and a range of stakeholders are interested in a more careful characterization of the change in number and intensity of such extreme weather patterns in the coming decades. 

In order to address these important questions, climate scientists routinely configure and run high-fidelity simulations under a range of different climate change scenarios. Each simulation produces 10s of TBs of high-fidelity output which requires automated analysis. Thus far, climate data analysts have relied entirely upon multi-variate threshold conditions for prescribing extreme weather patterns~\cite{prabhat:teca:2012}. Recent efforts~\cite{liu2016application,globenet, racah:nips17} have shown that deep learning can be successfully applied for detection, classification and localization of extreme weather patterns. In this paper, we push the frontier of deep learning methods to extract high-quality, pixel-level segmentation masks of weather patterns. 

%Analyzing extreme events in large datasets poses a significant challenge in climate science research. Conventional tools to analyze extreme events are  built upon human expertise, and they require subjective thresholds of relevant physical variables to define specific events. Tropical Cyclones (TCs), Extra Tropical Cyclones (ETCs), and Atmospheric Rivers (ARs) are important and impactful extreme weather events. Current methods to detect storms rely on sequential processing of the same data to detect each class of storm (TCs, ETCs, ARs, etc.). It would be significantly more efficient to detect all types of extreme weather events based on features/patterns that exist in multivariate climate datasets. Deep learning methods could achieve this goal when they are applied to physical variables such as integrated water vapor, surface pressure, and wind speed. Furthermore, traditional detection methods resort to subjective, arguably arbitrary thresholds, which may change with global warming. Accurate, efficient, and automatic tracking of extreme events can play a critical role in weather prediction if the network can learn precursors to these events. Deep neural networks may serve as an automated detector and tracker of extreme weather that relies on spatiotemporal patterns, not thresholds, in climate model simulations. With this tool, scientists can better study the environmental drivers that control the frequency, intensity, and location of extreme weather events and how they may change in a warming world. 

In this work, we use the TensorFlow~\cite{tensorflow2015-whitepaper,tensorflow-site} deep learning framework, which allows the programmatic definition of
even very complicated network {\em graphs} in tens of lines of Python code.  TensorFlow provides portability with its capability to map a
graph onto multi- and many-core CPUs as well as GPUs. Due to the heavy use of linear algebra-based primitives (e.g. convolutions), most
networks (including ours) perform very well on GPUs. The graph also captures the parallelism available in the computation, and TensorFlow
uses a dynamic scheduler to select which operation (or {\em layer}) to compute based on the availability of inputs.  (Scheduling is
performed independently on each process in a distributed job, leading to challenges with collective communication described in
Section~\ref{subsec:allreduce}.)

A deep learning model is trained by comparing its output to known {\em labels}, using a {\em loss function} to quantify the differences
between the two.  The model parameters (e.g. convolution weights) form a very-high-dimensional (typically millions) space, and training
becomes an optimization problem to find the point in the space that minimizes the loss.  Each layer is differentiable by construction,
and variants of gradient descent are typically used for optimization.  Since training sets can be very large, the most common variant is
stochastic gradient descent, which performs updates based on randomly-selected subsets (or {\em batches}) of the overall training set.

To parallelize training, a common technique is to replicate the model across ranks, with each rank processing a different local batch of images. Updates to the model are aggregated between ranks during each training step. As a deep learning problem is ``scaled out,'' the size of the global batch (combined batch across all ranks) grows but the size of the overall training set remains fixed,
making it distinct from traditional weak scaling.  The effect of batch size on convergence rate is not fully understood, but with the
right {\em hyperparameters} (parameters for the optimizer rather than the model), larger batches require fewer steps to converge, improving
overall time to solution if the parallel efficiency is sufficiently high.
Although an analogous form of strong scaling also exists (keeping the global batch size constant as worker count grows), it is generally only of interest when effective hyperparameters cannot be found for a larger global batch size.
%An analogous form of strong scaling also exists, where the global batch size is kept constant with increasing process count, resulting in reduced local batch sizes.   

\subsubsection{Segmentation Architecture}
\begin{figure}
\centering
\includegraphics[width=\linewidth]{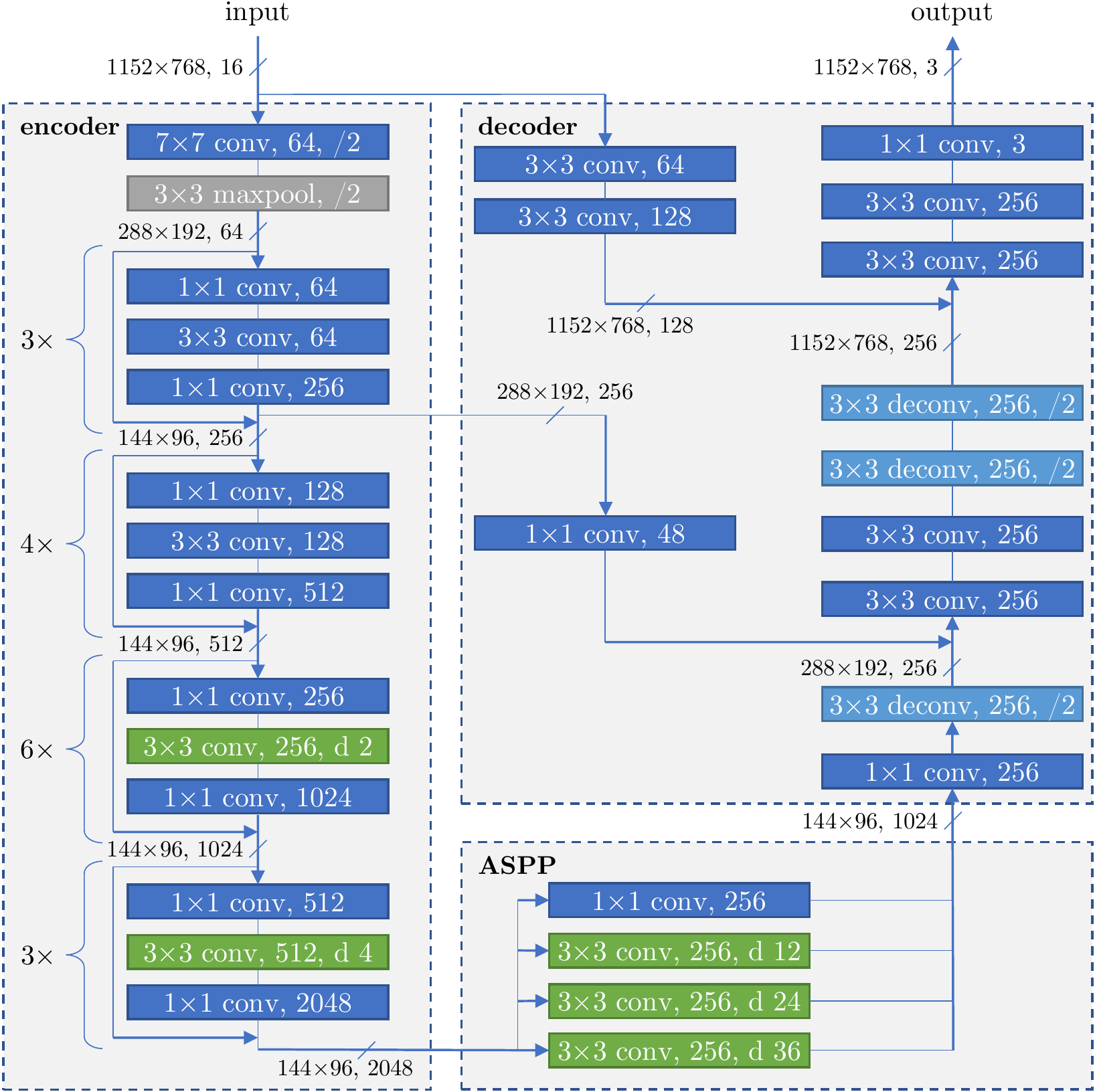}
\caption{\label{fig:deeplab}
Schematic of the modified DeepLabv3+ network used in this work.  The encoder (which uses a ResNet-50 core) and atrous spatial pyramid pooling (ASPP) blocks are changed for the larger input resolution.  The DeepLabv3+ decoder has been replaced with one that operates at full resolution to produce precise segmentation boundaries.  Standard convolutions are in dark blue, and deconvolutional layers are light blue.  Atrous convolution layers are in green and specify the
{\em dilation} parameter used.}
\vspace{-5mm}
\end{figure}

%\begin{figure}
%\centering
%\includegraphics[width=0.8\linewidth]{figures/Tiramisu-schematic.png}
%\caption{\label{fig:tiramisu} \textcolor{red}{@Sean: please replace by DeepLabv3+ figure.} Schematic of the Tiramisu architecture. It consists of \emph{Dense blocks} (blue) which consist of a sequence of full convolutions (conv) (stride=1) followed by a non-linearity.  Dense blocks are followed by a \emph{transition down} (pink) that uses convolutions with stride=2 to shrinks the number of feature dimensions, creating a bottle neck. A similar up-sampling path is shown on the right hand side of the figure. Skip connections (dotted lines) between the two paths ensure better reconstructions. The figure only shows two blocks while the model we employed had five blocks. Orange circles in each path represent concatenations of features, improving training of deep networks.}
%\vspace*{-3mm}
%\end{figure}

High-level frameworks like TensorFlow make it convenient to experiment with different networks.  We evaluated two
very different networks for our segmentation needs.  The first is a modification of the \emph{Tiramisu} network~\cite{jegou2017one}.
%Fig.~\ref{fig:tiramisu} shows a diagram of a simple Tiramisu network with three dense blocks, comprised of a sequence of convolutional layers.
Tiramisu is an extension to the Residual Network (ResNet) architecture~\cite{he2016deep} which introduced \emph{skip connections} between layers to force the network to learn residual corrections to layer inputs. Where ResNet uses addition, Tiramisu uses concatenation to combine the inputs of one or more layers of the network with their respective outputs. The Tiramisu network is comprised of a \emph{down path} that creates an \emph{information bottle-neck} and an \emph{up path} that reconstructs the input. To perform pixel-level segmentation, Tiramisu includes skip connections spanning the down and up paths to allow re-introduction of information lost in the down path.
%help regularize the up path and generate better images by forcing the loss for each layer to be consistent. 
%\emph{Skip connections} between the down and up paths help regularize the up path to 
%generate labels that match the input data.
Our Tiramisu network uses five dense blocks in each
direction, with 2,2,2,4 and 5 layers respectively (top to bottom). We then train the model using adaptive moment estimation (ADAM)~\cite{kingma2014adam}. 

The second network we evaluated is based on the recent DeepLabv3+ network \cite{deeplabv3plus2018} and is shown in
Figure~\ref{fig:deeplab}.  DeepLabv3+ is an
{\em encoder-decoder architure} that uses well-proven networks (in our case ResNet-50) as a core.  The {\em encoder}
performs a function similar to Tiramisu's down path but avoids loss of information by replacing some of the downscaling
with {\em atrous convolution}.  Atrous convolutions sample the input sparsely according to a specified {\em dilation}
factor to detect larger features.  This simplifies the {\em decoder} (corresponding to Tiramisu's up path) considerably. 
Our modifications to these existing networks are described in Section~\ref{sec:mod_arch}.
%The atrous
%convolutions result in a more computationally expensive network and the standard DeepLabv3+ design makes the compromise
%of performing segmentation at one-quarter resolution (i.e. $288 \times 192$ rather than $1152 \times 768$) to keep
%the computation tractable for less-powerful systems.  Our use case demands sharper boundaries, but we have the
%unparalleled performance of Summit available, so we replace the standard DeepLabv3+ decoder with one that operates
%at full resolution.

\subsubsection{Climate Dataset and Ground Truth Labels}
We utilize 0.25-degree Community Atmosphere Model (CAM5) output for this study. Climate variables are stored on an $1152\!\times\! 768$ spatial grid, with a temporal resolution of 3 hours. Over 100 years of simulation output are available in HDF5 files. Ideally, climate scientists would hand-label pixel masks corresponding to events.
In practice, scientists currently use a combination of heuristics to produce masks on large datasets.
%In lieu of the availability of a large scale dataset, we use a combination of heuristics to produce ground truth data.
The first step is to process climate model output with the Toolkit for Extreme Climate Analysis~\cite{prabhat:teca:2012,prabhat:teca:2015} to identify TCs. A floodfill algorithm is used to create spatial masks of ARs~\cite{shields2018atmospheric}, which provides the labels for our training process.

There are about 63K high-resolution samples in total, which are split into 80\% training, 10\% test and 10\% validation sets. We use all available 16 variables (water vapor, wind, precipitation, temperature, pressure, etc). The pixel mask labels correspond to 3 classes: Tropical Cyclone (TC), Atmospheric River (AR) and background (BG) class.

\subsection{Contributions}
Motivated by the problem of finding extreme weather patterns in climate data, our paper makes the following contributions:
\begin{itemize}
\item{We adapt state-of-the art Tiramisu and DeepLabv3+ architectures to solve segmentation problems on high resolution, multi-variate scientific datasets.}
\item{We make a number of system-level innovations in data staging, efficient parallel I/O, and optimized networking collectives to enable our DL applications to scale to the largest GPU-based HPC systems in the world (Section~\ref{subsec:system-innovations}).}
\item{We make a number of algorithmic innovations to enable DL networks to converge at scale (Section~\ref{subsec:dl-innovations}).}
\item{We demonstrate good scaling on up to \herogpucount\, GPUs, obtaining \halfprecperfsust\, PF/s sustained performance and a parallel efficiency of \halfprecefficiency\% (Section~\ref{sec:results}) for half precision. The peak performance we obtained at that concurrency and precision was \halfprecperfpeakef\, EF/s.}
\item{Our code is implemented in TensorFlow and Horovod; our performance optimizations are broadly applicable to the general deep learning + HPC community, our stack is already being used by several other projects.}
\end{itemize}

While our work is conducted in the context of a specific science driver, most of our proposed innovations are applicable to generic deep learning workloads at scale.
\section{State of the Art}
\label{sec:intro}

%\textcolor{red}{ quantitative discussion of current SoA, with emphasis on performance-related aspects  (1 p max)}
%\textcolor{blue}{Sean, Josh: please update/review}

\subsection{State-of-the-art in Scientific Deep Learning}

In recent years, the scientific community has begun to adopt deep learning methods and frameworks as tools for scientific analysis and discovery~\cite{oreilly,radovic:2018,Mathuriya:2018,Guzik:2016,DanielG:2018,regier:2015}. Early applications were focused on adapting off-the-shelf convolutional neural networks from natural image processing applications or recurrent neural networks from speech recognition applications (for a review see~\cite{karpatne_climate_2017}). There is currently a shift in the community towards incorporating scientific principles (e.g. physical laws such as energy or momentum conservation) and common assumptions 
(e.g. temporal and/or spatial coherence). Some recent examples in the domain areas related to ours include simulation of local wind field patterns via coupled autoencoder architectures~\cite{henningh_latnet_2017},  turbulence modeling for climate simulations via deep networks trained with loss functions that incorporate physical terms~\cite{wang_piml_2017}, and supervised applications of extreme weather pattern detection~\cite{liu2016application}. The field of physics-informed deep learning for scientific and engineering applications is in its infancy, and this paper is a timely contribution focused on exploring the computational limits of representative architectures that many of the above approaches are based on. 

\subsection{State-of-the-art in Large-Scale Deep Learning}

Modern-day deep neural networks build upon the work laid out by McCullogh and Pitt~\cite{mcculloch1943logical}, and Rosenblatt (perceptron)~\cite{rosenblatt1958perceptron}. While forming the foundation for
deep learning, these early models often struggled as the network size increases, limiting their utility in the analysis of
complex systems.  More recently, work by Krizhevsky~\cite{krizhevsky2012imagenet} opened the flood gates for modern day Deep Learning, showing impressive performance on hard vision tasks using large supervised deep networks.  This breakthrough was
made possible in part by the rapid increase in computational power of modern computing systems. Since then, the complexity of tasks and the size of the networks have been growing steadily over the years, arguably requiring larger and more powerful platforms. 

There has been more recent work on scaling deep learning up to larger node counts and performance.  Preferred Networks, Inc. demonstrated ResNet-50 converging to 75\% accuracy in 15 minutes using the ChainerMN~\cite{chainermn_mlsys2017} framework on 1024 NVIDIA Tesla P100 GPUs at a total global batch size of 32K for 90 epochs~\cite{pfn-rn50}.  Jia at al.~\cite{Tencent}, concurrent with this work, demonstrated scaling to 2048 NVIDIA Tesla P40 GPUs at 64K batch size, achieving convergence in 6.6 minutes using TensorFlow.  To effectively utilize leadership class systems, we need to push scaling significantly further than previous work.
Most work on classification networks uses relatively small images from the computer vision community.  Our
work extends Deep Learning to handle much larger input in the form of snapshots from a scientific simulation.  These 
"images" can be millions of pixels in size and generally have many more channels than the red, green, and blue of
commodity imaging sensors.
%Also, our application domain and input data is also different than well studied classification networks \textcolor{red}{(JRD: this sentence may need rewording)}.
We are also contending with a significantly larger dataset that pushes the limits of the file system and requires new data handling techniques.

\section{Innovations}
\label{sec:innovations}

%\textcolor{red}{What are the innovations and how were they achieved, 2 pages max}
%\textcolor{blue}{Sean, Josh, Mike Houston: please update/review}
\subsection{System Innovations}
\label{subsec:system-innovations}

\subsubsection{High speed parallel data staging}
\label{subsec:data-staging}
Modern neural networks require large amounts of input data and therefore training can easily be bottlenecked by an inability to bring
input data to the GPU in a timely fashion.
%\textcolor{red}{Do we have updated data for DLV3?} [SJT] Tiramisu is worse for I/O, so I think we can leave this untouched.
For instance, a single GPU training our modified Tiramisu network can consume 
189 MB/s, already above the capabilities of a local hard drive, which means the 6 GPUs on a Summit node require a combined 1.14 GB/s. A training run using 1024 nodes therefore requires a sustained read bandwidth of 1.16 TB/s, and running on the full Summit system will require 5.23 TB/s, more than twice the target performance of the GPFS file system.

Summit makes available 800 GB of high-speed SSD storage on each node to help with local bandwidth needs.  %Although this is non-volatile storage, it is scrubbed for each job.  %Applications must stage data at the start of a job and time spent on this staging is computing cycles wasted.  
While a training data set can be quite large (the
climate data used in this study is currently 3.5 TB), in a distributed training setting, it suffices for each node to have access to a significant fraction of the overall data set. The images selected by each rank are combined to form a batch, so a sufficient (and independently selected) set of samples for each rank to choose from results in batches that are statistically very similar to a batch selected from the entire data set. In our experiments, 250 images per GPU (1500 per node) are sufficient to maintain convergence.

Unfortunately, a naive staging script that asked each of 1024 nodes to copy its own subset of the full data set from GPFS
required 10-20 minutes to complete and rendered the global file system nearly unusable for other users of the machine during that time.  With this approach, each individual file from the data set was being read by 23 nodes on average. To address this, we developed a distributed data staging system that first divides the data set into disjoint pieces to be read by each rank.  Each rank's  I/O throughput was further improved by running multiple threads that perform file reads in parallel -- using eight threads instead of one increased the achieved read bandwidth from 1.79 GB/s on average to 11.98 GB/s, an improvement of $6.7\times$. Once all files in the data set have been read from GPFS, point-to-point MPI messages are used to distribute copies of each file to other nodes that require it. This approach takes advantage of the significantly higher
bandwidth of the Infiniband network and places no further load on the file system. Our improved script is able to stage in data for 1024 (4500) nodes on Summit in under 3 (7) minutes.

On Piz Daint, where no local SSDs are available, the only
node local storage with sufficient bandwidth to feed the P100 GPU is the Linux \texttt{tmpfs}
(DRAM), which has much more limited capacity. 
%\textcolor{red}{To address this, our staging script modifies the data as it is copied
%from the global file system, removing input channels that will not be used by the current network and optionally
%reducing the precision of the input data if the network is computing in FP16 rather than FP32.  (While these data
%reductions could have been performed on the data set in global storage, it would require multiple copies of the data set to be maintained, one for each set of input channels and numerical precision.)  Even for FP32 data, this reduced the 
%locally staged data to 15 GB, a much more manageable fraction of DRAM. An additional benefit of this data reduction is to further improve the performance of the distribution phase of the staging script, as the distribution over MPI is done after the reduction is performed.}

\subsubsection{Optimized data ingestion pipeline}

Although the staging of input data into fast local storage eliminates bottlenecks and variability from global file system reads, optimization is also required for the TensorFlow input pipeline that reads the input files and converts them into the tensors that are fed through the network.  By default, the operations to read and transform input data are placed in the same operation graph as the networks themselves, causing idle time on the GPU while the CPU performs input-related tasks.  This serialization can be eliminated by enabling the prefetching option of TensorFlow datasets, which allows the input pipeline to run ahead of rest of the network, placing processed input data into a queue.  As long as the queue remains non-empty, the network can obtain its next input immediately upon completion of the previous one.  The queue depth can be made deep enough to insulate against variability in the input processing rate, but the average
production rate must still exceed the average consumption
rate. As a further optimization, TensorFlow allows for concurrent processing of multiple input files using its \texttt{map} operator; however, the HDF5 library used to read the climate data serializes all operations, negating the benefit of parallel operation.  By using the Python
\texttt{multiprocessing} module, we were able to transform these parallel worker threads into parallel worker processes, each
using its own instance of the HDF5 library.  With 4 background processes taking care of reading and processing input data, the input pipeline can more closely match the training throughput of both networks, even when using FP16 precision.

\subsubsection{Hierarchical all-reduce}
\label{subsec:allreduce}

Network training is distributed across multiple GPUs using Horovod~\cite{sergeev2018horovod}.
Horovod is a Python module that uses MPI to transform a single-process TensorFlow application into a data-parallel
implementation.  Each MPI rank creates its own identical copy of the TensorFlow operation graph.  Horovod then inserts all-reduce operations into the back-propagation computation to average the computed gradients from each rank's network.  Ranks update their local models independently, but (assuming consistent initialization) the use of gradients averaged across all the ranks results in identical updates (i.e. synchronous distributed training).
Although it is possible for a TensorFlow+Horovod implementation to use multiple GPUs per rank, we adopted the simpler approach of using a different MPI rank for each GPU (i.e. 6 ranks per node on Summit), allowing the same code to be used on both Summit and Piz Daint.  Horovod has been shown to have good scalability up to 1024 GPUs, but as we scaled further, we saw a dramatic loss in parallel efficiency resulting from two issues.

The first issue was a bottleneck on the first rank, which acts as a centralized scheduler for Horovod operations.  As each TensorFlow process is independently scheduling the operations in its graph, different ranks might attempt to execute their 
all-reduce operations in different orders, resulting in deadlock. Horovod resolves this by dynamically reordering all-reduce operations to be consistent across all ranks. Each rank sends a message to the controller (rank 0) indicating readiness
to perform a given all-reduce operation.  Once the controller has received messages from all ranks for one or more operations, it sends a return message
to every rank with an ordered list of tensors on which to perform collective operations.  Our network has
over a hundred allreduce operations per step, forcing the controller to receive and then send millions of messages per
second for larger jobs.  A distribution of the scheduling load is not possible, as all ranks must agree on a total order of
collective
operations to perform, so we chose instead to perform hierarchical aggregation of the control messages.  The ranks are
organized into a tree of configurable radix $r$, and each node in the tree waits for readiness messages from all of its
direct children (and its own local operation) before sending a readiness message to its parent in the tree.  Rank 0 sits
at the root of the tree and uses the original Horovod algorithm for scheduling, but operates as if there were only $r{+}1$
ranks to
coordinate.  When a rank receives a message to start collective operations, it first relays that message to its children (if any) and then initiates the collective.  This recursive broadcast approach guarantees that no rank sends or receives more
than $r{+}1$ messages for each tensor, reducing the message load to mere thousands of messages per second, regardless of scale.
Tuning of broadcast tree shapes can be important when latency is a concern, but TensorFlow's 
dynamic scheduler makes it fairly tolerant to small latency differences, and we observed no
measureable performance difference for values of $r$ between 2 and 8.

The second issue to address was the performance of the collective all-reduce operations themselves.  The existing Horovod
implementation is able to reduce data residing on GPUs in two different ways, either by a standard \texttt{MPI\_Allreduce}
or by using the NVIDIA Collective Communications Library (NCCL)\cite{nccl-site}.  Both have their strengths: MPI often uses
tree-based communication patterns for performance at scale, while NCCL uses a systolic ring approach that takes
advantage of the bandwidth of GPUs that are connected with NVLink within a Summit node.  To obtain both the scalability of MPI
and the local bandwidth improvements of NCCL, we implemented a hybrid all-reduce approach.  Data is first reduced across the GPUs
within a node using NCCL.  Once those 6 ranks have the same locally-reduced data, 4 of the ranks (two on each CPU
socket) each perform an
\texttt{MPI\_Allreduce} on a quarter of the data, sharing with the corresponding rank on every other node and obtaining their
quarter of the final result.  Finally, NCCL broadcast operations are used within the node to ensure each of the 6 GPUs has
a full copy of the entire all-reduce result.  The decision to have 4 local ranks perform MPI operations was based on 
experimentation, but suggests that a 1:1 mapping between communicating processes and virtual network devices is the most efficient strategy
on Summit (each node has a dual-rail Mellanox IB ConnectX-5 EX adapter that is virtualized as 4 IB devices). With only a single GPU per node, Piz Daint does not benefit from this hybrid all-reduce implementation, but with
the trend towards higher GPU counts per node, we expect this optimization to be beneficial on future machines as well.

\subsection{Deep Learning Innovations}
\label{subsec:dl-innovations}
\subsubsection{Weighted loss}
The image segmentation task for climate analysis is challenging because of the high class imbalance: about 98.2\% of the
pixels are BG and about 1.7\% of the overall pixels are ARs. Pixels labelled as TCs make up less than 0.1\% of the total.
With an unweighted loss function, each pixel contributes equally to the loss function, and a network can (and did, in practice) achieve high
accuracy (98.2\% in our case) by simply predicting the dominant background class for all pixels.  To improve upon this situation, we
use a weighted loss calculation in which the loss for each pixel is weighted based on its
labeled class.  The per-pixel weight map is
calculated as part of the input processing pipeline and provided to the GPU along with the input image.
Our initial experiments used the inverse of the class frequencies for weights, attempting to
equalize the collective loss contribution from each class.  We found that this approach led to
numerical stability issues, especially with FP16 training, due to the large difference in
per-pixel loss magnitudes.  We examined more moderate weightings of the classes and found that
using the inverse square root of the frequencies addressed stability concerns while still 
encouraging the network to learn to recognize the minority classes (see Figure~\ref{fig:deeplab-seg}).

\subsubsection{LARC}
Layer-wise adaptive rate control (LARC)~\cite{larc-paper} is designed to control the magnitude of weight updates by keeping them small compared to the norm of layer's weights. LARC uses a separate independent learning rate
for every layer instead of every weight. The magnitude of the update
is defined with respect to the weight's norm. LARC improves the accuracy of large networks, especially when trained using large batch sizes. Compared to layer-wise adaptive rate scaling (LARS)~\cite{2017arXiv170803888Y}, LARC removes the need for complex learning rate warm-up techniques and is thus much easier to use. Given all these advantages, we use LARC for the results reported in this study. 

\subsubsection{Multi-channel segmentation}
Traditional image segmentation tasks work on 3-channel RGB images. However, scientific datasets can be comprised of many channels: in case of the CAM5 climate dataset, those can incorporate fields such as temperature, wind speeds, pressure values, and humidity at different altitudes.  Our initial
experiments on Piz Daint used 4 channels that were thought to be the most important, but when
the network was moved to Summit, the additional computational capabilities allowed the use of all
16 channels, which improved the accuracy of the models dramatically.  The optimal subset of channels to
use likely lies in between these two, and we plan to take advantage of the ability to rapidly
train this network at scale to tune for the right subset.

\subsubsection{Gradient lag} \label{subsec:gradient_lag}
Most of the all-reduce operations required for gradient computation can be overlapped with other computation, but the
top-most layer's gradient computation is a sequential bottleneck for a standard optimizer.  The network-induced latency of
this computation can limit performance at large scale.  To improve parallel efficiency, we modified the optimizer to use
the gradients computed in the previous step when performing weight updates.  In addition to improving the overlap of
communication and computation, this {\em lagging} of the gradients allows Horovod to more efficiently batch the tensors for all-reduce computations, 
increasing network throughput.  Although a change to the optimizer usually requires changes to the hyperparameters to
maintain convergence properties, the performance benefit is usually worth the effort at large scale.  A similar gradient lagging strategy, known as elastic averaging SGD (EASGD) was shown to be effective, with even larger degrees of lag~\cite{EASGD}.

\subsubsection{Modifications to the neural network architectures} \label{sec:mod_arch}
The developers of the original Tiramisu network advocate the use of many layers with a relatively small growth rate per layer (e.g. 12
or 16)~\cite{jegou2017one} and our initial network design used a growth rate of 16.  This network learned well, but
performance analysis of the resulting
TensorFlow operations on Pascal and Volta GPUs found considerable room for improvement and we determined that a growth rate of
32 would be significantly more efficient.  To keep the overall network size roughly the same, we reduced the number of layers
in each dense block by a factor of two and changed the convolutions from $3\! \times\!3$ to $5\!\times\!5$ to maintain the same
receptive field.  Not only was the new network much faster to compute, we found that it trained faster and yielded a better
model than our original network.

For DeepLabv3+, the atrous
convolutions result in a more computationally expensive network than Tiramisu. The standard DeepLabv3+ design makes the compromise
of performing segmentation at one-quarter resolution (i.e. $288 \times 192$ rather than $1152 \times 768$) to keep
the computation tractable for less-powerful systems, at the cost of fidelity in the resulting masks. The irregular and fine-scale nature of our segmentation labels requires operating at the native resolution of the dataset. With the
unparalleled performance of Summit available for this work, we were able to replace the standard DeepLabv3+ decoder with one that operates at full resolution, thereby benefiting the science use case.

\section{Performance Measurement}
\label{sec:measurement}

\begin{figure*}[ht!]
\centering
\tiny
\resizebox{\textwidth}{!}{%
\renewcommand{\arraystretch}{1.1}%
\begin{tabular}{|lcccccc|} \hline
\multirow{2}{*}{\bf Network} & \bf Operation Count & \multirow{2}{*}{\bf GPU Model (System)} & \multirow{2}{*}{\bf Precision} & \bf Training Rate & \bf Performance & \bf \%  \\ 
& \bf (TF/sample) & & & \bf (samples/s) & \bf (TF/s) & \bf Peak \\ \hline

\multirow{2}{*}{DeepLabv3+} & \multirow{2}{*}{14.41} &  \multirow{2}{*}{V100 (Summit)} & FP16 & 2.67 & 38.45 & 31 \\
&&& FP32 & 0.87 & 12.53 & 80  \\ \hline
\multirow{3}{*}{Tiramisu} & \multirow{2}{*}{4.188} &  \multirow{2}{*}{V100 (Summit)} & FP16 & 5.00 & 20.93 & 17 \\
&&& FP32 & 1.91 & 8.00 & 51  \\
& 3.703* & P100 (Piz Daint) & FP32 & 1.20 & 4.44 & 48 \\ \hline
\end{tabular}
}
\caption{\label{fig:single-gpu-perf} Single GPU performance results from training the Tiramisu and DeepLabv3+ networks.  Results are shown for all tested systems using FP32 and FP16 precision where relevant. Note that the operation count for Tiramisu on Piz Daint (marked with an asterisk) is computed from a modified network using 4 out of the 16 available input data channels. }
\end{figure*}

%\textcolor{red}{what application was used to measure performance (1pg). System and environment where performance was measured (1pg).}
%\textcolor{blue}{Mass: please review}

Training performance of a deep neural network is generally reported in images (or batches) per
second, but it can be useful to
convert these numbers into floating point performance (i.e. FLOP/s).  To do so, we incorporate some
Python code that performs an analysis on the TensorFlow operation graph constructed by the
application.  The nodes of the graph are traversed and the number of FLOPs required for each operation
is computed.  This graph-based analysis is essential for computing an accurate FLOP count
when working with an application that defines multiple networks that share nodes.

For convolution nodes, additional
analysis was required as there are multiple algorithmic formulations available, some of which
require different quantities of floating point operations. TensorFlow dynamically tunes the algorithm choice
for best performance, so it was necessary to use the API tracing capability in cuDNN to determine the algorithm selection. With the current versions of TensorFlow and cuDNN, we found that all convolutions were performed using either implicit GEMMs or direct convolutions.
For example, a $3\!\times\!3$ direct convolution on a $1152 \!\times\!768$ image
with $48$ input channels, $32$ output channels and a batch size of $2$ requires
$3 \!\times \!3 \!\times\! 1152 \!\times \!768 \!\times\! 48 \!\times\! 32 \!\times\! 2\! \times\! 2 = 48.9\! \times 10^9$
FLOPs. (The final factor of 2 follows the normal convention of counting both multiplies and
additions as FLOPs.)

Once the FLOP count per step has been determined, we normalize this by the number of samples (images) per step. Based on the number of steps we can then compute the number of samples processed in that step per rank and compute statistics on the time series of steps. 
If not otherwise stated, we compute the mean number of processed samples for every step over ranks and the median of the result over time and quote this as our sustained throughput. We further compute an (asymmetric) error bar based on the central 68\% confidence interval (computed from the 0.16 and 0.84 percentiles) over time. Using the FLOP per sample we can then compute a FLOP rate by multiplying the total processed samples per second with the FLOP per sample.

As is common for deep learning training situations, a series of additional calculations is carried out on the validation data set after each epoch, i.e. a full pass over the training data has been performed.  Our data staging technique holds the number of steps in an epoch constant as we scale to larger node counts, keeping the epoch sizes large enough that this overhead is
negligible once amortized over the steps.

\begin{figure*}[t!]
\small
\resizebox{\textwidth}{!}{%
\renewcommand{\arraystretch}{1.1}%
\begin{tabular}{|l@{ }l@{ }l|rrrr|rrrr|rrrr|rrrr|}
\hline
& & & 
& \multicolumn{6}{c}{\bf Tiramisu} & &
& \multicolumn{6}{c}{\bf DeepLabv3+} & \\
& & &
\multicolumn{4}{c|}{\bf FP32 Training} & \multicolumn{4}{c|}{\bf FP16 Training} &
\multicolumn{4}{c|}{\bf FP32 Training} & \multicolumn{4}{c|}{\bf FP16 Training} \\
{\bf Category} & & & 
\begin{tabular}{@{}c@{}}{\bf \#} \\ {\bf Kern}\end{tabular} &
\begin{tabular}{@{}c@{}}{\bf \%} \\ {\bf Time}\end{tabular} &
\begin{tabular}{@{}c@{}}{\bf \%} \\ {\bf Math}\end{tabular} &
\begin{tabular}{@{}c@{}}{\bf \%} \\ {\bf Mem}\end{tabular} &
\begin{tabular}{@{}c@{}}{\bf \#} \\ {\bf Kern}\end{tabular} &
\begin{tabular}{@{}c@{}}{\bf \%} \\ {\bf Time}\end{tabular} &
\begin{tabular}{@{}c@{}}{\bf \%} \\ {\bf Math}\end{tabular} &
\begin{tabular}{@{}c@{}}{\bf \%} \\ {\bf Mem}\end{tabular} &
\begin{tabular}{@{}c@{}}{\bf \#} \\ {\bf Kern}\end{tabular} &
\begin{tabular}{@{}c@{}}{\bf \%} \\ {\bf Time}\end{tabular} &
\begin{tabular}{@{}c@{}}{\bf \%} \\ {\bf Math}\end{tabular} &
\begin{tabular}{@{}c@{}}{\bf \%} \\ {\bf Mem}\end{tabular} &
\begin{tabular}{@{}c@{}}{\bf \#} \\ {\bf Kern}\end{tabular} &
\begin{tabular}{@{}c@{}}{\bf \%} \\ {\bf Time}\end{tabular} &
\begin{tabular}{@{}c@{}}{\bf \%} \\ {\bf Math}\end{tabular} &
\begin{tabular}{@{}c@{}}{\bf \%} \\ {\bf Mem}\end{tabular}
\\ \hline
\multirow{2}{*}{Forward} & \multirow{2}{*}{$\Big\lbrace$} & Convolutions &
71 & 31.4 & 51.7 & 64.4 &
95 & 25.3 & 21.2 & 101.2 &
239 & 33.3 & 75.6 & 21.2 &
158 & 18.1 & 52.0 & 20.7 \\
& & Point-wise &
563 & 7.9 & * & 82.1 &
564 & 12.2 & * & 76.8 &
870 & 3.2 & * & 73.2 &
829 & 6.4 & * & 51.6 \\
\multirow{2}{*}{Backward} & \multirow{2}{*}{$\Big\lbrace$} & Convolutions &
95 & 49.2 & 65.7 & 62.9 &
113 & 38.3 & 28.0 & 66.7 &
127 & 49.0 & 102.7 & 9.0 &
195 & 36.7 & 51.2 & 18.7 \\
& & Point-wise &
113 & 0.7 & * & 59.6 &
123 & 2.8 & * & 47.9 &
145 & 0.9 & * & 44.9 &
157 & 3.1 & * & 27.3 \\
{Optimizer} & & & 
1056 & 0.5 & * & 25.9 &
1056 & 0.7 & * & 33.3 &
1219 & 0.3 & * & 30.6 &
1219 & 0.5 & * & 31.3 \\
\multicolumn{3}{|l|}{Copies/Transposes} &
%{Copies / Transposes} & & & 
388 & 5.5 & & 78.0 &
530 & 12.3 & & 60.8 &
535 & 8.6 & & 66.9 &
708 & 26.1 & & 48.3 \\
\multicolumn{3}{|l|}{Allreduce (NCCL)} &
%{Allreduce (NCCL)} & & & 
25 & 5.1 & * & 1.6 &
30 & 5.4 & * & 3.5 &
35 & 4.6 & * & 1.2 &
30 & 7.2 & * & 1.1 \\
\multicolumn{3}{|l|}{Type Conversions} &
%{Type Conversions} & & &
&&&&
143 & 0.1 & & 22.2 &
&&&&
201 & 0.2 & & 51.3 \\
GPU Idle & &&
&&&&
    & 2.9 & & &
&&&&
    & 1.7 & & \\
%Idle Time & & & & ??.? & - & - \\
\hline
Total     & & &
2311 & & 48.5 & 62.3 &
2654 & & 16.1 & 69.8 &
3170 & & 75.5 & 20.2 &
3497 & & 28.2 & 27.7
\\ \hline
\end{tabular}%
}

\caption{\label{fig:single-node-perf}%
Summary of single node performance analysis of training for both Tiramisu (left) and Deeplabv3+ (right) networks. Kernels are grouped by category, and results are shown for both FP32 and FP16 training.  The fraction of time spent in kernels from each category is shown along with the fraction of peak math and memory performance achieved by kernels in that category.  An asterisk (*) is used to indicate values less than $0.1\%$.  Values reported are subject to some measurement uncertainty (see text).}
\end{figure*}

\subsection{HPC Systems and Environment}

\subsubsection{Piz Daint}
\label{sec:piz}
Piz Daint at CSCS~\cite{pizdaint-site} is a hybrid Cray XC40/XC50 system. We will only consider the XC50 portion of the machine in this paper. The latter is comprised of 5320 hybrid CPU+GPU nodes. The CPU are single-socket Intel Xeon E5-2695v3 with 12 hardware cores which can host 2 HyperThreads each at 2.6 GHz. Each node has 64 GB of DDR memory and is further equipped with one NVIDIA Pascal GPU (P100) with 16 GB HBM2 memory and 32 GB/s PCIe bidirectional bandwidth. The nodes are connected by a low-latency high-bandwidth Aries interconnect with a diameter-5 Dragonfly topology.  The peak single-precision floating point performance of the machine is 50.6 PF/s, twice the quoted 25.3 PF/s double-precision performance~\cite{top500-pizdaint-site}. The global LUSTRE file system offers a peak bandwidth of 744 GB/s for reads and a total capacity of 28 PB.

\emph{Software environment:}
On Piz Daint, we use TensorFlow v1.6, compiled with CUDA 8.0 and the cuDNN 7.1.1 backend. We use our improved Horovod with hierarchical control plane which is based on the official v0.12.0 release. We compile it against Cray MPICH v7.6.0 and enable CUDA-aware collectives.

\subsubsection{Summit}
\label{sec:summit} 
Summit is the new leadership class supercomputer installed at the Oak Ridge National Laboratory (ORNL). This system is the current top-ranked supercomputer in the TOP500 rankings, the first on the list to surpass the 100 double-precision PF mark on the HPL benchmark~\cite{top500-summit-site}.  
%\textcolor{red}{In its final configuration, the system will have approximately 4600 nodes, each one with two IBM Power9 CPUs and 6 NVIDIA Volta 16GB GPUs.  At the time of writing, around 4500 of these nodes are available to Early Science projects.} 
The system is comprised of 4608 nodes, each equipped with two IBM Power 9 CPUs and 6 NVIDIA Volta GPUs (V100) with 16 GB HBM2 memory.
Each Power 9 CPU is connected to 3 Volta GPUs using NVIDIA high-speed interconnect NVLink, capable of  300 GB/s bi-directional bandwidth.
Each node has 512 GB of system memory and a 1.6 TB NVMe disk, half of which is available to jobs to be used as burst buffer.
Dual-rail EDR Infiniband cards connect all the nodes using a non-blocking fat-tree topology.
The nodes can access a POSIX-based IBM Spectrum Scale parallel file system with a current capacity of 3 PB and approximate maximum speed of 30 GB/s. 

The Volta architecture includes Tensor Cores that provide mixed-precision operations. 
In each cycle, each of the 640 Tensor Cores can perform 64 floating-point Fused-Multiply-Add (FMA) operations  with  input values in half precision and output values either in half (FP16) or single precision (FP32).
Deep Learning workloads are able to use mixed-precision. Utilizing the Tensor Cores, each Volta GPU can perform 125 trillion floating-point operations per second, resulting in a peak node performance of 750 TF/s.

\emph{Software environment:}
On Summit, we use TensorFlow v1.8 compiled with CUDA 9.2 and cuDNN v7.2 backend. We again use our improved Horovod with hierarchical control plane which is based on the official v0.12.0 release. We compile it against IBM Spectrum MPI v10.2 and also NCCL v2.2.13 for fast GPU-based intranode all-reduces.

\section{Performance Results}
\label{sec:results}

%\textcolor{red}{include scalability (weak and strong), time to solution, efficiency (of bottleneck resources), and peak performance (2 pgs max).}
%\textcolor{blue}{Sean, Josh, Thorsten, Mass, Mike Houston: please update/review}

\subsection{Single GPU Performance}\label{sec:single-node-performance}
%\textcolor{red}{Need DLV3 discussion added. Josh: Changed up this section. Needs review.}

%\textcolor{red}{Using the methodology described in Section~\ref{sec:measurement}, we determined that processing
%a single image using the full set of 16 input channels on Summit requires 4.188 TF
%(regardless of precision), while the reduced network run on Piz Daint requires 3.703 TF.
%Due to GPU memory constraints, a single image is processed at a time when using FP32 precision.
%The reduced memory footprint of the FP16 network allows batches of two images to be processed per
%GPU.
%On Piz Daint, each GPU is able to process 1.20 images per second, for a sustained single-GPU performance of 4.44 TF/s, roughly 48\% of the theoretical peak performance of the P100 GPU.
%The full network on each GPU in Summit processes 1.91 images per second with FP32 precision, achieving
%8.00 TF/s of sustained performance, about 51\% of the V100 GPU's theoretical peak.  When using
%FP16 precision, the performance on Summit jumps to 5.00 images per second, or 20.93 TF/s, representing 17\% of the peak performance of the V100's Tensor Cores.}

Using the methodology described in Section~\ref{sec:measurement}, we determined the number of floating point operations required to process a single image with the Tiramisu and DeepLabv3+ networks. Combining these values with the sustained training rate (in samples/s) per GPU yields the sustained single GPU compute performance in Flop/s for each network. For both networks, a single image per GPU is processed per training step when FP32 precision is used, while for FP16, the lower memory footprint enables batches of two images per GPU to be processed during each training step. Single GPU performance results from this analysis can be found in Figure~\ref{fig:single-gpu-perf}. From the tabulated data, we observed that the DeepLabv3+ network utilizes compute resources more efficiently than Tiramisu, achieving a higher percentage of peak Flop/s for both FP32 and FP16 computations. However, when comparing FP32 to FP16 computations across all results, the FP16 results are notably less efficient.

To determine the source of these performance inefficiencies, we made a detailed analysis of the 
work performed on the GPU using the CUDA profiling tools.  Figure~\ref{fig:single-node-perf}
provides a
summary of this analysis, with further per-network details shown in Figure~\ref{fig:single-node-perf-tiramisu} (Tiramisu)
and Figure~\ref{fig:single-node-perf-deeplab} (DeepLabv3+).  In order to capture the cost of all-reduce operations, this
analysis was performed on a job running across 4 Summit nodes (24 GPUs), so the numbers differ slightly from the
single GPU performance discussed above.  Multiple runs were required to measure different performance counters, and the
non-determinism in TensorFlow's execution of the graph adds some noise to the measurements (which, in some cases, causes ratios to slightly exceed 100\%).
Further, each training step requires thousands of kernels, making a
traditional roofline analysis difficult.  Instead, we grouped kernels into eight categories and
looked at the computational and memory needs of each category.  All of the computationally intensive kernels are in the forward and backwards convolutions, which get
fairly good utilization of the FP32 computing resources.  However, the analysis shows that the Tiramisu network's
convolution kernels become memory limited when using FP16 precision.
This is a fundamental limitation of the Tiramisu-style
network due to its small filter sizes per layer.
The convolutions in the DeepLabv3+ use much larger channel counts per layer, resulting in higher computational
intensity.  This reduces the overall memory demand and improves datapath utilization.
In addition to the convolutional layers, neural networks require many point-wise operations in
the forward and backward passes (e.g. bias and dropout layers) as well as in the optimizer.
The most expensive of these are in the forward pass, and get very good memory utilization for
both FP32 and FP16 precisions. A small but significant contribution to the overall step time comes
from copies and transpose operations that are inserted by TensorFlow. Although both
networks can be implemented without extra copies (by assembling layers in place), the TensorFlow
graph optimization pass is not able to make such a specific optimization. As a final optimization, we modified the data layout of the decoder stage of the DeepLapv3+ network to produce fewer extraneous transposes. This modification yielded a ~10\% speedup compared to the original code for our largest scale run.
Finally, the NCCL
kernels used for the intra-node portion of the all-reduce operations are bottlenecked by the 
bandwidth of the NVLink connections rather than the DRAM bandwidth (as described in 
Section~\ref{subsec:allreduce}, the MPI portion of the all-reduces is performed on the CPU
concurrently with GPU and is not shown here).

Our analysis found that the GPU is kept completely busy for the FP32 cases, indicating that any
performance improvements have to come from optimizing or eliminating some of the kernels running
on the GPU. The most beneficial kernels to optimize are the convolutions, but with so many
different kernels being used, the effort would be significant, and would deny the application the
benefit of any improvements that are made to the cuDNN library.  For example, a move from cuDNN v7.0.5 to
v7.1.2 early in the project resulted in a 5\% performance improvement with no changes to the application.
We explored a move
away from TensorFlow, implementing the network directly with cuDNN library calls, but the resulting
code was much harder to maintain than the TensorFlow version.  A 5-10\% performance gain was not
worth the impact on programmer productivity.
%The vast improvement in programmer
%productivity was deemed to be worth a 5\% reduction in runtime performance.
The final optimization
strategy, and the one we are pursuing, is to make incremental improvements within TensorFlow to
improve the memory management and fuse some of the point-wise operations together to reduce the
number of times tensors are read and written to DRAM.  This might also allow the batch size to be increased, which would also improve the efficiency of the convolutional stages.

With the use of significantly faster math in the FP16 cases, the memory-bound kernels consume a
larger fraction of the overall step time, and any optimizations to eliminate copies or fuse 
point-wise tasks will help the FP16 even more than FP32.  The profile for the FP16 also shows some periods where the GPU has run out of work, suggesting that code running on the CPU such as the input pipeline or the TensorFlow scheduler may require additional optimization
as well.

\subsection{Scaling Experiments}
\label{sec:scaling-experiments}
We perform several scaling experiments on Piz Daint and Summit. On Piz Daint, we ran the Tiramisu network only, while on Summit, both Tiramisu and DeepLabv3+ networks were run. The experiment setup is slightly different for the two systems and we explain the details below. We bind one MPI rank to each GPU which amounts to one rank per node on Piz Daint and six ranks per node on Summit.

%\subsubsection{Tiramisu}

On Piz Daint, we scale the training up from a single GPU to the full machine, i.e. 5300 nodes. We also compare the scaling behavior when staging input data against reading it from the global Lustre file system.
On Summit, we run with a single GPU as a baseline, but then sweep from 1 to \heronodecount\, nodes using all
6 GPUs per node (i.e. 6 to \herogpucount\, GPUs).

\begin{figure*}[ht]
\centering
\begin{minipage}{.494\textwidth}
\includegraphics[width=0.99\linewidth]
{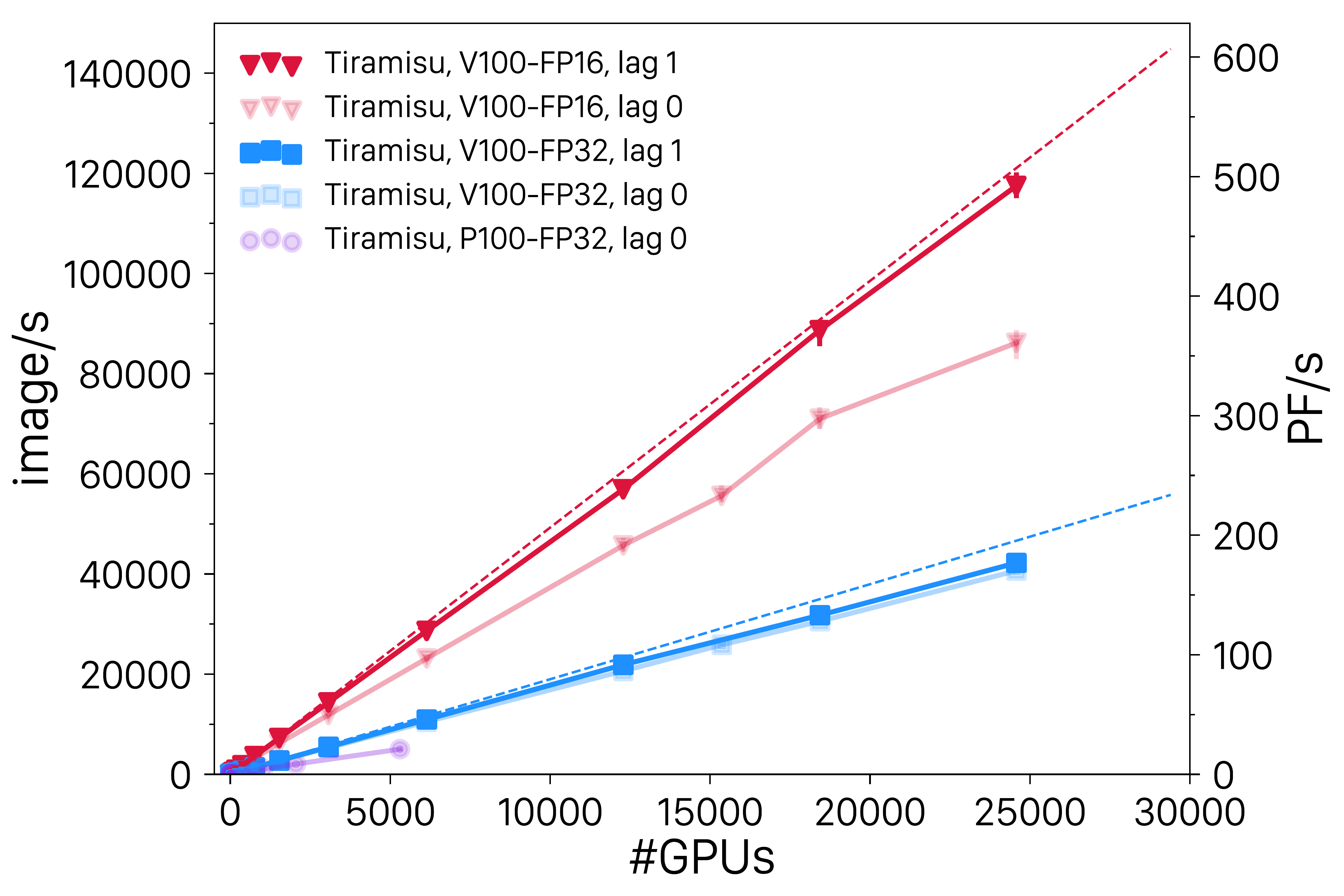}
\subcaption{Tiramisu}
\end{minipage}
\begin{minipage}{.494\textwidth}
\includegraphics[width=0.99\linewidth]
{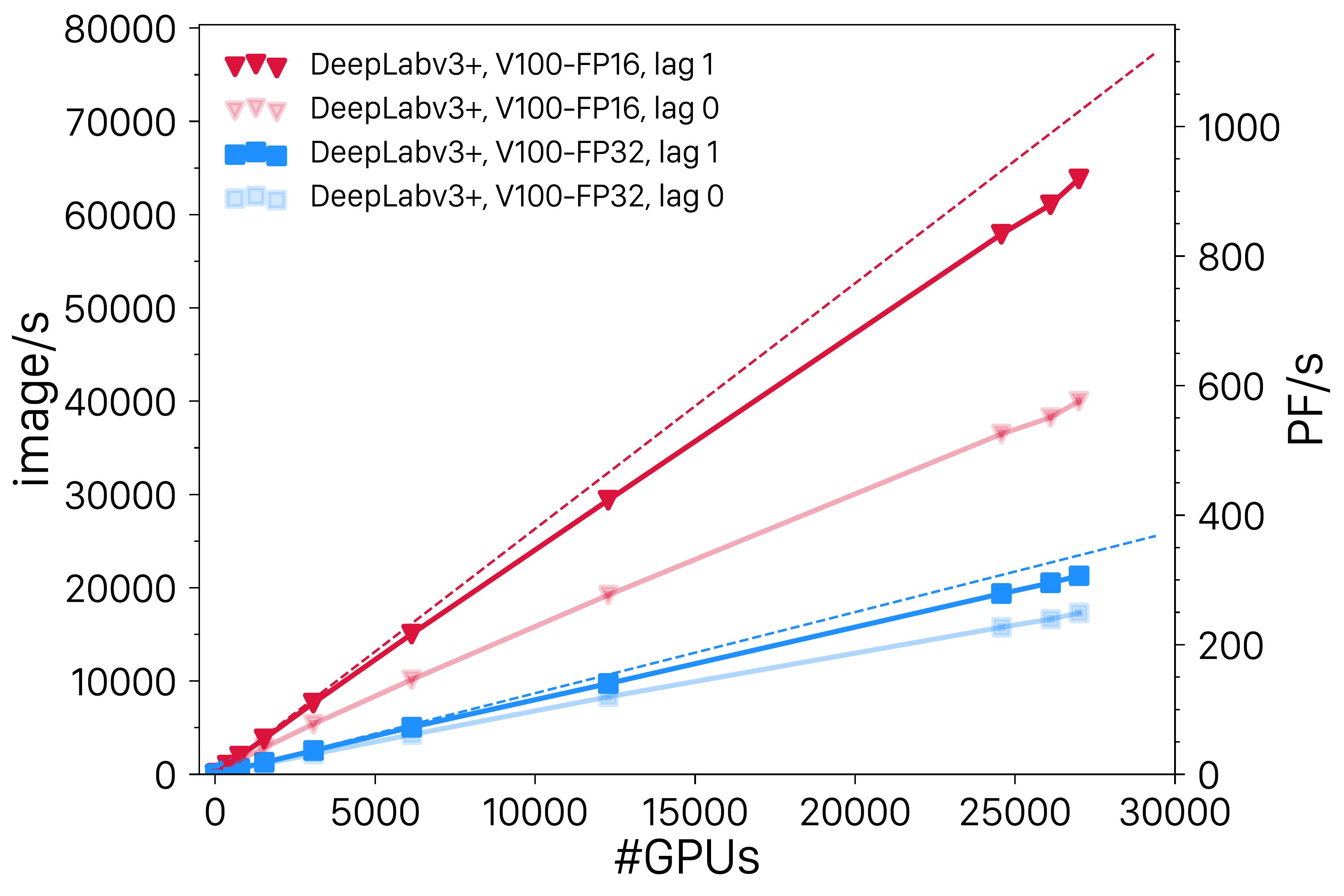}
\subcaption{DeepLabv3+}
\end{minipage}
\caption{\label{fig:weakscale_tiramisu}Weak scaling results in terms of images/sec and sustained performance in PF/s on Summit (FP16 and FP32, Tiramisu and DeepLabv3+) and Piz Daint (FP32, Tiramisu). The dashed lines represent the ideal scaling lines for the different architectures and precisions.}
\end{figure*}

The scaling results are shown in Figure~\ref{fig:weakscale_tiramisu}. We find that the training performance of Tiramisu scales to a sustained 21.0 PF/s on the full Piz Daint machine, achieving parallel efficiencies of 83.4\% at 2048 nodes and 79.0\% at 5300 nodes in FP32. On Summit, scaling Tiramisu to 4096 nodes yields a sustained throughput of 176.8 PF/s and 492.2 PF/s for FP32 and FP16 precision respectively, maintaining parallel efficiencies above 90\% in both cases.
Moving on to DeepLabv3+, scaling to \heronodecount\, nodes with FP32 precision yields a sustained throughput of \singleprecperfsust\, PF/s and a parallel efficiency of \singleprecefficiency\%.
The FP16 network reaches a peak \halfprecperfpeakef\, EF/s, sustained \halfprecperfsust\, PF/s and \halfprecefficiency\% parallel efficiency at that scale. The highest performing results were obtained on Summit in the cases with gradient lag (see Section~\ref{subsec:gradient_lag}) enabled, corresponding to the data labeled ``lag 1" in Figure~\ref{fig:weakscale_tiramisu}. The results clearly indicate the effectiveness of the lagged scheme in improving the overall application scalability.

%The results show that this technique is very effective at improving scalability in the cases it was utilized. 

%\begin{figure*}
%\centering
%\includegraphics[scale=0.15]{figures/weak_scaling_tiramisu.pdf}
%\caption{Weak scaling results for Tiramisu on Piz Daint (FP32, left panel), Summit FP32 (center) and Summit FP16 (right panel).}
%\end{figure*}

\begin{figure}[ht]
\centering
\includegraphics[width=0.9\linewidth]{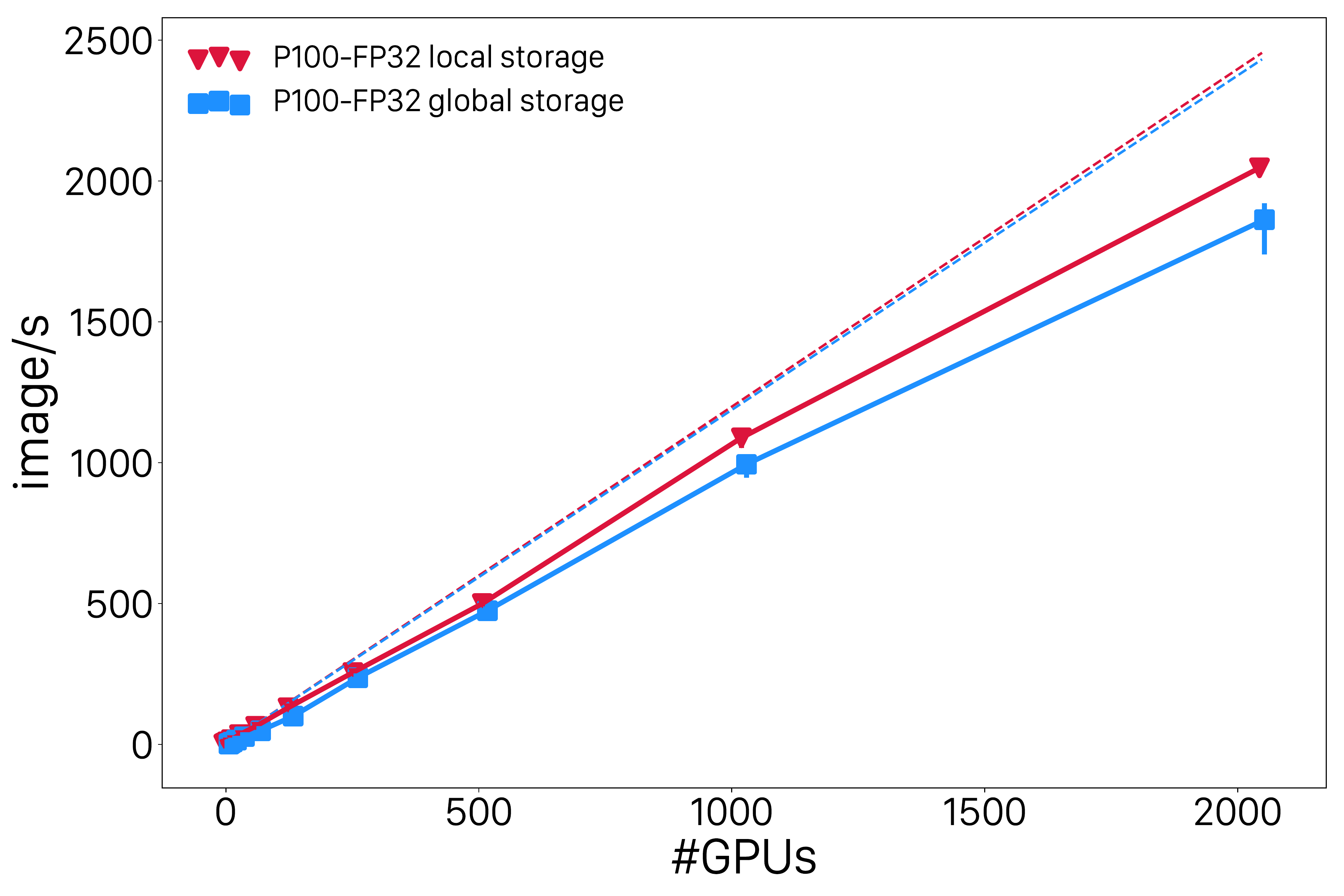}
\caption{\label{fig:fsdep_tiramisu}Dependence of weak scaling on input data location on Piz Daint.}
\end{figure}

To demonstrate the benefit of the data staging process described in
Section~\ref{subsec:data-staging}, we experimented on Piz Daint with reading input data directly
from the global file system and highlight results in Figure~\ref{fig:fsdep_tiramisu}.
Performance matches the runs using data staging at lower node counts, but the difference becomes apparent at larger scales.
On 2048 GPUs, the parallel efficiency has dropped to 75.8\%, a 9.5\% penalty for not staging the input data in the local
\texttt{tmpfs} storage. Additionally, the throughput shows larger variability. At this scale, the neural network is demanding nearly 110 GB/s of input data, very close to the file system's limit of 112 GB/s. Therefore, we did not attempt to scale beyond 2048 nodes without data staging.

\subsection{Convergence at Scale}

A major challenge for deep neural networks is to maintain convergence properties (and with good
accuracy) as the network training is scaled out. To demonstrate the stability of our network at large 
scales, we performed longer runs on up to 1024 Summit nodes using both the FP32 and FP16
precision modes, training the network to convergence. As with the scaling runs, the dataset is
resampled to put 1500 files per node, improving the statistical properties of the large batches
being used at this scale. The training in each case was performed for a fixed number of epochs (targeting a total training time of just over two hours). 

\begin{figure}[t]
\centering
\includegraphics[width=0.9\linewidth]{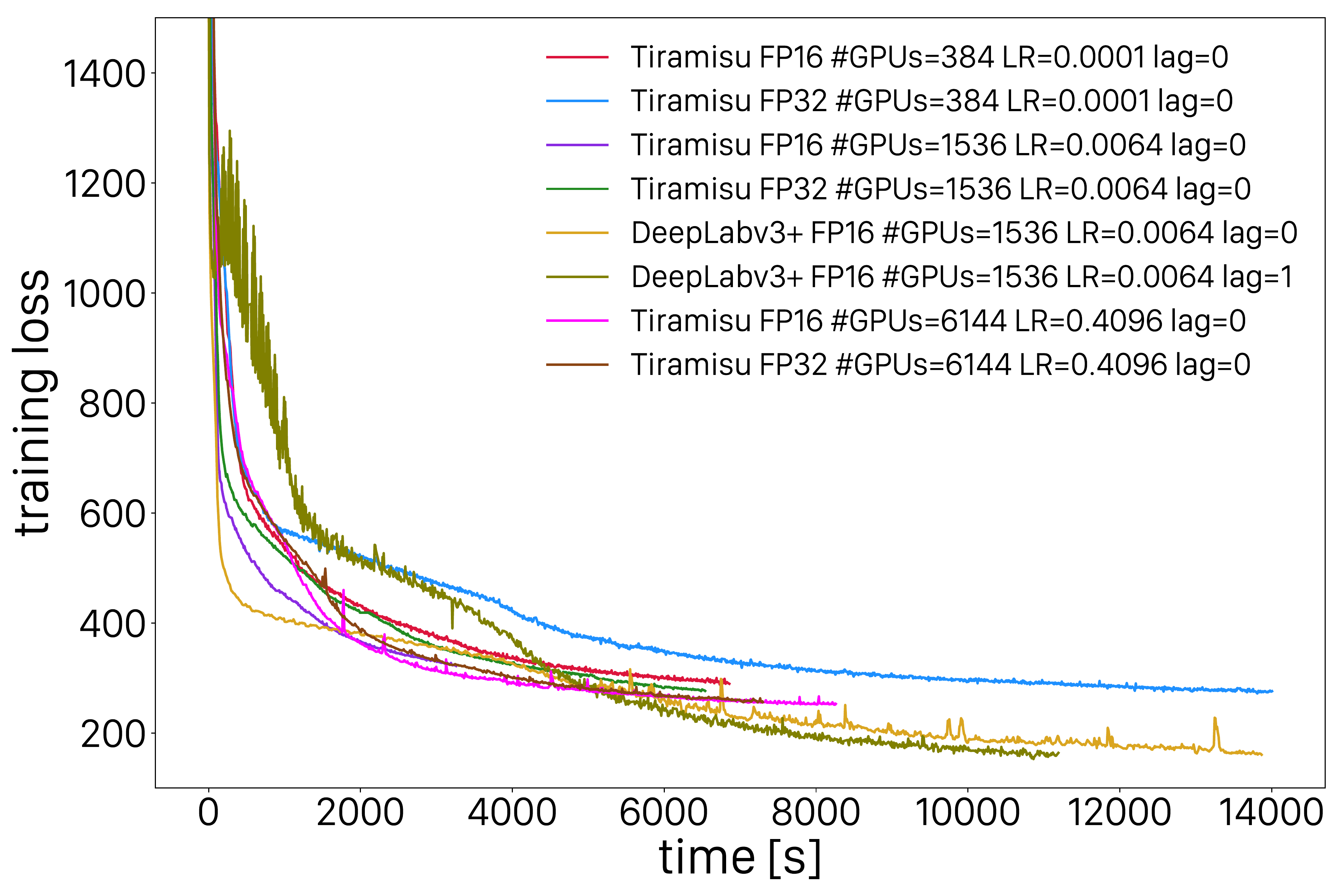}
\caption{\label{fig:hero_loss} Training loss curves for various concurrencies and precisions for the Tiramisu and DeepLabv3+ architectures.}
\end{figure}

The training loss curve for these runs are shown in Figure~\ref{fig:hero_loss} along with curves
for runs at smaller scales (384 GPUs and 1536 GPUs).  Moving averages over 10 step windows are
used to filter out step-to-step fluctuations in the loss. As can be seen in Figure~\ref{fig:hero_loss}, all of the configurations are converging with both FP16 and FP32. Tiramisu as well as DeepLabv3+ network is
stable at large scale with the initially chosen set of hyperparameters. Tuning of
hyperparameters is always necessary when scaling up a network, and we expect that the time to solution will improve further as they are dialed in. There are a few other important things to notice in Figure~\ref{fig:hero_loss} 1) FP16 converges in significantly less time that FP32; 2) DeepLabV3+ generally converges faster than Tiramisu; 3) And lag0 vs lag1 with DeepLabV3+ has nearly identical training loss curves. The ability to perform these experiments in an hour or two rather than days is a key enabler to being able to perform training at these scales and explore the hyperparameter and algorithm space.

%\textcolor{red}{\subsubsection{Convergence}
%The training loss curve is displayed in Fig.~\ref{fig:hero_loss}. We measured the training loss (displayed in red) as rolling epoch average. To illustrate the convergence we additionally applied a moving average with window size of 800 steps, which roughly corresponds to one epoch at that scale. The validation loss and intersection-over-union score (IoU) are displayed in blue and purple respectively.
%The plot shows the network is still converging and thus a longer training run should have improved the IoU score even further.
%The plot also suggests that the learning rate we used might be on the higher end so a slightly reduced learning rate might also lead to better accuracy but at the cost of a longer wall clock time.}
%\begin{figure}
%\centering
%\includegraphics[scale=0.21]{figures/hero_loss_curve.pdf}
%\caption{\label{fig:hero_loss} Moving mean averaged loss curve obtained on 5300 GPUs on Piz Daint.}
%\end{figure}

%\subsection{HERO Run}

%For our HERO run, we perform a calculation on 2560 Summit nodes (15360 GPUs) in FP16 and FP32 precision. As in the convergence test we discussed before, we stage 1500 files per node into NVM.
%We compute the FLOP rate as described in Section~\ref{sec:measurement}. The main difference to the above sections is that, since this is a realistic training run, we include the epoch metric calculation overhead as well in our final performance estimates. This will only impact the reported sustained performance and not the peak performance but it is expected to be negligible for the stated precision as this overhead is paid on only 0.1\% of the total steps.
%\begin{itemize}
%\item\emph{Peak Flop Rate}: we achieve a peak throughput of 26385.2 samples/s which corresponds to a \textbf{peak performance of 110.5 PetaFLOP/s FP32 (single precision)}. For FP16, we achieve a peak throughput of 62841.0 samples/s which yields a \textbf{peak performance of 263.2 PetaFLOPS/s}. This  corresponds to 45.8\% and 13.7\% of the theoretical peak respectively.
%\item\emph{Sustained Flop Rate}: we achieve a sustained throughput of 25848.0 samples/s which amounts to a \textbf{sustained single precision FP32 performance of 108.3 PetaFLOP/s}. For FP16, we obtain a sustained throughput of 55610.0 samples/s. This corresponds to a \textbf{sustained Flop Rate of 232.9 PetaFLOPS/s}.
%\end{itemize}

\subsection{Climate Science Results}
\label{sec:science_results}

\begin{figure*}[ht]
\begin{subfigure}[b]{0.64\textwidth}
\includegraphics[width=\textwidth]{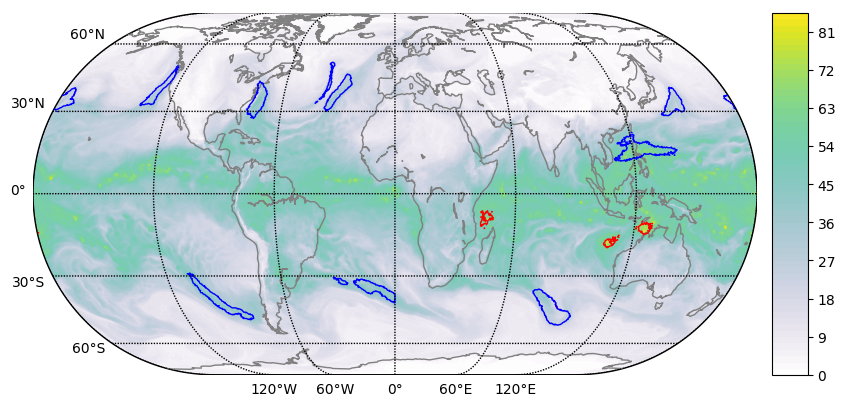}
\caption{Segmentation masks overlaid on a globe.  Colors (white->yellow) indicate IWV (integrated water vapor, $kg/m^{2}$), one of the 16 input channels used by the network.
\label{subfig:deeplab-seg-full}}
\end{subfigure}
\hfill
\begin{subfigure}[b]{0.33\textwidth}
\includegraphics[width=\textwidth]{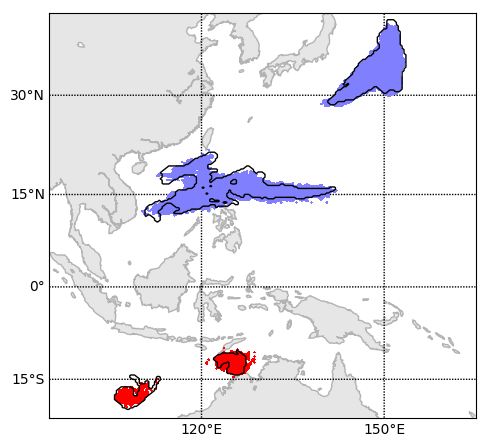}
\caption{Detailed inset showing predictions (red and blue) vs. labels used in training (black).
\label{subfig:deeplab-seg-zoom}}
\end{subfigure}
\caption{Segmentation results from modified DeepLabv3+ network.  Atmospheric rivers (ARs) are labeled in blue, while
tropical cyclones (TCs) are labeled in red.
\label{fig:deeplab-seg}}
\vspace{-4mm}
\end{figure*}

%\begin{figure}[t]
%\centering
%\includegraphics[scale=0.2]{figures/test_combined_epoch50_estep0_rank0.png}
%\caption{Ground truth data showing background (in black), TCs (in gray) and ARs (in white). False positives by the model are shown in cyan (ARs) or pink (TCs), while false negatives appear in red (ARs) or green (TCs).  Blue and yellow indicate confusion between ARs and TCs by the model.}
%\label{fig:tiramisu-seg}
%\end{figure}

Segmentation accuracy is often measured using the {\em intersection over union} (IoU) metric.  The Tiramisu network obtained an IoU of 59\% on our validation data set, while our modified DeepLabv3+ network was able to achieve 73\% IoU.
%The mean intersection over union (IoU) \textcolor{red}{(JOSH: We don't define IoU in this paper. Do we need it?)} obtained by the network was about 59\% for the model.
Visually, this translates into qualitatively pleasing masks as seen in Figure~\ref{fig:deeplab-seg}. Not only does
the network find the same atmospheric features, it makes a very good approximation of their exact boundaries.  In some cases, the boundaries predicted by the model appear to be superior to the labels provided by heuristics. One of the
tropical cyclones in Figure~\ref{subfig:deeplab-seg-zoom} does suffer from overprediction.  This is an expected consequence of our weighted loss function, which penalizes a false negative on a TC by roughly $37 \times$ more than a false positive.

\section{Implications}
\label{sec:future}

%\textcolor{red}{Implications for future systems and applications (1pg max)}
\subsection{Climate Science}

This is the first successful demonstration of the application of Deep Learning for extracting pixel-level segmentation masks in the climate science community. This analysis opens the door for more sophisticated characterization of extreme weather than what has been possible before. Prior to this work, climate scientists reported coarse summary statistics such as number of global storms. In contrast, we can now compute conditional precipitation, wind velocity profiles and power dissipation indices for \emph{individual} storm systems. These sophisticated metrics will enable us to characterize the \emph{impact} from each event (physical damage to infrastructure, flooding, monetary loss, etc) with unprecedented fidelity. 

In the future, we will explore advanced architectures that can consider temporal evolution of storms. This will increase the resident size of the network architecture, requiring model (as well as data) parallelism for efficient execution. We also plan on working with the climate science community to generate high quality ground truth datasets without resorting to heuristics. Developing accessible interfaces for specifying masks, and bootstrapping the process using online, semi-supervised methods is an area for further investigation.

\subsection{Future Systems}
%\textcolor{blue}{Mike, Sean: we need you to take a pass here.}
Scaling Deep Learning further on future exascale machines with purely data parallel techniques will prove to be numerically difficult. Techniques such as LARC have increased the total global batch size that can converge, but we view the incorporation of model parallel approaches as being indispensable in the foreseeable future. Systems like Summit (with high speed NVLink connections between processors) are amenable to domain decomposition techniques that split layers across processors. Exploring model parallel implementation across nodes is a natural extension, that will require investments in more complex collectives in software libraries like Horovod and NCCL, and optimizations at the algorithm as well at network switch level. 
%\textcolor{red}{This should provide a factor of 6 more total scaling at the same global batch size on Summit.}  
%Beyond this, we will need to explore model parallel techniques across multiple nodes. This will require investments in more complex collectives in software libraries like Horovod and NCCL, and optimizations at the algorithm and network switch level.

Additional training performance optimizations will increase the rate at which we need to feed input data to the networks, further exacerbating the parallel I/O problem. While compression techniques can be used at the expense of already heavily utilized main processors, more memory close to the compute elements, such as node-local non-volatile memory, may help reduce the pressure on the global file system. There is also a potential for processing at the storage layer itself to aid in data processing and augmentation. Generally the stress on the dataplane and communication layers will quickly increase, requiring holistic approaches towards hardware and software co-design. 

To conclude, we believe that field of Deep Learning is poised to have a major impact on the scientific world. Scientific data requires training and inference at scale, and while Deep Learning might appear to be a natural fit for existing petascale and future exascale HPC systems, careful consideration must be given towards balancing various subsystems (CPUs, GPUs/accelerators, memory, storage, I/O and network) to obtain high scaling efficiencies. Sustained investments are required in the software ecosystem to seamlessly utilize algorithmic innovations in this exciting, dynamic area.

%strongly believe that while Deep Learning is a perfect use case for HPC platforms, sustained investment is required in the software and hardware ecosystem in order to maximally 

%and signficant invesment in both, to keep up with the rapid pace of deep learning algorithm innovation and performance improvements in GPUs and similar processors with respect to deep learning performance.

\section{Conclusions}
\label{sec:conclusions}

We have presented the first exascale-class deep learning application. Motivated by the important problem of segmenting extreme weather patterns, we have successfully applied the Tiramisu and DeepLabv3+ architectures to high resolution, multi-variate climate datasets. We developed a number of enhancements to the deep learning algorithms (custom loss function, optimization schemes, channels and network architecture) to obtain excellent qualitative and quantitative results. We built upon a number of system-level optimizations (advanced data staging, optimized data ingestion, and hierarchical all-reduce communication) to scale the scientific application to an unprecedented level of concurrency (\heronodecount\, Summit nodes, \herogpucount\, Volta GPUs), scaling efficiency (\halfprecefficiency\%) and performance (\halfprecperfpeakef\, EF/s peak, \halfprecperfsust\, PF/s sustained). Our work extends open-source TensorFlow and Horovod tools, thereby benefiting the broader scientific and commercial deep learning communities.  The environment we have developed is already in use by other teams on Summit and the methodologies will extend to current and future HPC platforms.

\section*{Acknowledgments}
This research used resources of the National Energy Research Scientific Computing Center (NERSC), a DOE Office of Science User Facility supported by the Office of Science of the U.S. Department of Energy under Contract No. DE-AC02-05CH11231. This work was supported by a grant from the Swiss National Supercomputing Centre (CSCS) under Project ID g107.
We thank Nicholas Cardo, Andreas Joksch, Miguel Gila and the CSCS staff for assistance in using Piz Daint. We thank Paul Tucker and Rajat Monga from Google for helpful discussions pertaining to TensorFlow. Michael Wehner, Karthik Kashinath, Burlen Loring, Travis O'Brien and Bill Collins from LBNL were instrumental in motivating the climate science problem and providing datasets. 
This research used the Summit system at the Oak Ridge Leadership Computing Facility at the Oak Ridge National Laboratory, which is supported by the Office of Science of the U.S. Department of Energy under Contract No. DE-AC05-00OR22725.
%Our results were obtained using the Summit supercomputer at Oak Ridge Leadership Computing Facility (OLCF), a DOE Office of Science user facility at  Oak Ridge National Laboratory (ORNL). 
We are very grateful to OLCF staff: Veronica Melesse Vergara; Don Maxwell, and Matthew Ezell for their assistance with the runs, and Arjun Shankar; Ashley Barker; Tjerk Straatsma and Jack Wells for programmatic support.

\bibliographystyle{IEEEtran}
\bibliography{reference}

\clearpage
\onecolumn
\begin{multicols}{2}
\appendix
\end{multicols}
%\section{Detailed Performance Analysis Results}
\begin{figure}[h]
\small
\resizebox{\textwidth}{!}{%
\renewcommand{\arraystretch}{1.1}%
\begin{tabular}{|l@{ }l@{ }l|rrrrrrr|rrrrrrr|}
\hline
& & & 
& \multicolumn{5}{c}{\bf FP32 Training} & & 
& \multicolumn{5}{c}{\bf FP16 Training} & \\
{\bf Category} & & & 
\begin{tabular}{@{}c@{}}{\bf \#} \\ {\bf Kern}\end{tabular} &
\begin{tabular}{@{}c@{}}{\bf Time} \\ {\bf (ms)}\end{tabular} &
\begin{tabular}{@{}c@{}}{\bf Math} \\ {\bf (TF)}\end{tabular} &
\begin{tabular}{@{}c@{}}{\bf Mem} \\ {\bf (GB)}\end{tabular} &
\begin{tabular}{@{}c@{}}{\bf \%} \\ {\bf Time}\end{tabular} &
\begin{tabular}{@{}c@{}}{\bf \%} \\ {\bf Math}\end{tabular} &
\begin{tabular}{@{}c@{}}{\bf \%} \\ {\bf Mem}\end{tabular} &
\begin{tabular}{@{}c@{}}{\bf \#} \\ {\bf Kern}\end{tabular} &
\begin{tabular}{@{}c@{}}{\bf Time} \\ {\bf (ms)}\end{tabular} &
\begin{tabular}{@{}c@{}}{\bf Math} \\ {\bf (TF)}\end{tabular} &
\begin{tabular}{@{}c@{}}{\bf Mem} \\ {\bf (GB)}\end{tabular} &
\begin{tabular}{@{}c@{}}{\bf \%} \\ {\bf Time}\end{tabular} &
\begin{tabular}{@{}c@{}}{\bf \%} \\ {\bf Math}\end{tabular} &
\begin{tabular}{@{}c@{}}{\bf \%} \\ {\bf Mem}\end{tabular}
\\ \hline
\multirow{2}{*}{Forward} & \multirow{2}{*}{$\Big\lbrace$} & Convolutions &
71 & 172.4 & 1.40 & 100.0 & 31.4 & 51.7 & 64.4 &
95 & 105.5 & 2.79 & 96.1 & 25.3 & 21.2 & 101.2 \\
& & Point-wise &
563 & 43.6 & $< 0.1$ & 32.2 & 7.9 & & 82.1 &
564 & 51.1 & $< 0.1$ & 35.3 & 12.2 & & 76.8 \\
\multirow{2}{*}{Backward} & \multirow{2}{*}{$\Big\lbrace$} & Convolutions &
95 & 270.5 & 2.79 & 153.2 & 49.2 & 65.7 & 62.9 &
113 & 159.7 & 5.58 & 95.8 & 38.3 & 28.0 & 66.7 \\
& & Point-wise &
113 & 4.1 & $< 0.1$ & 2.2 & 0.7 & & 59.6 &
123 & 11.6 & $< 0.1$ & 5.0 & 2.8 & & 47.9 \\
{Optimizer} & & & 
1056 & 3.0 & $< 0.1$ & 0.7 & 0.5 & & 25.9 &
1056 & 3.0 & $< 0.1$ & 0.9 & 0.7 & & 33.3 \\
{Copies / Transposes} & & & 
388 & 30.5 & - & 19.8 & 5.5 & & 78.0 &
530 & 51.5 & - & 28.2 & 12.3 & & 60.8 \\
{Allreduce (NCCL)} & & & 
25 & 28.2 & $< 0.1$ & 0.4 & 5.1 & & 1.6 &
30 & 22.4 & $< 0.1$ & 0.7 & 5.4 & & 3.5 \\
{Type Conversions} & & &
&&&&&&&
143 & 0.5 & - & 0.1 & 0.1 & & 22.2 \\
GPU Idle & &&
&&&&&&&
    & 12.0 &  &     & 2.9 & & \\
%Idle Time & & & & ??.? & - & - \\
\hline
Total     & & &
2311 & 549.9 & 4.19 & 308.5 & & 48.5 & 62.3 &
2654 & 417.3 & 8.38 & 262.1 & & 16.1 & 69.8
\\ \hline
\end{tabular}%
}
%\captionsetup{width=\textwidth}
\caption{\label{fig:single-node-perf-tiramisu}%
Detailed single node performance analysis of Tiramisu network training for FP32 (left) and FP16 (right) precision.  Kernels are grouped by category, with the total time, FLOPs, and memory traffic reported for each.  The fraction of time spent in kernels from each category is shown along with the fraction of peak math and memory performance achieved by kernels in that category. Values reported are subject to some measurement uncertainty (see text).}
\end{figure}

\begin{figure}[h]
\small
\resizebox{\textwidth}{!}{%
\renewcommand{\arraystretch}{1.1}%
\begin{tabular}{|l@{ }l@{ }l|rrrrrrr|rrrrrrr|}
\hline
& & & 
& \multicolumn{5}{c}{\bf FP32 Training} & & 
& \multicolumn{5}{c}{\bf FP16 Training} & \\
{\bf Category} & & & 
\begin{tabular}{@{}c@{}}{\bf \#} \\ {\bf Kern}\end{tabular} &
\begin{tabular}{@{}c@{}}{\bf Time} \\ {\bf (ms)}\end{tabular} &
\begin{tabular}{@{}c@{}}{\bf Math} \\ {\bf (TF)}\end{tabular} &
\begin{tabular}{@{}c@{}}{\bf Mem} \\ {\bf (GB)}\end{tabular} &
\begin{tabular}{@{}c@{}}{\bf \%} \\ {\bf Time}\end{tabular} &
\begin{tabular}{@{}c@{}}{\bf \%} \\ {\bf Math}\end{tabular} &
\begin{tabular}{@{}c@{}}{\bf \%} \\ {\bf Mem}\end{tabular} &
\begin{tabular}{@{}c@{}}{\bf \#} \\ {\bf Kern}\end{tabular} &
\begin{tabular}{@{}c@{}}{\bf Time} \\ {\bf (ms)}\end{tabular} &
\begin{tabular}{@{}c@{}}{\bf Math} \\ {\bf (TF)}\end{tabular} &
\begin{tabular}{@{}c@{}}{\bf Mem} \\ {\bf (GB)}\end{tabular} &
\begin{tabular}{@{}c@{}}{\bf \%} \\ {\bf Time}\end{tabular} &
\begin{tabular}{@{}c@{}}{\bf \%} \\ {\bf Math}\end{tabular} &
\begin{tabular}{@{}c@{}}{\bf \%} \\ {\bf Mem}\end{tabular}
\\ \hline
\multirow{2}{*}{Forward} & \multirow{2}{*}{$\Big\lbrace$} & Convolutions &
239 & 404.4 & 4.80 & 77.1 & 33.3 & 75.6 & 21.2 &
158 & 147.9 & 9.61 & 27.6 & 18.1 & 52.0 & 20.7 \\
& & Point-wise &
870 & 39.3 & $< 0.1$ & 25.9 & 3.2 & & 73.2 &
829 & 52.3 & $< 0.1$ & 24.3 & 6.4 & & 51.6 \\
\multirow{2}{*}{Backward} & \multirow{2}{*}{$\Big\lbrace$} & Convolutions &
127 & 596.0 & 9.61 & 48.5 & 49.0 & 102.7 & 9.0 &
195 & 300.2 & 19.21 & 50.5 & 36.7 & 51.2 & 18.7 \\
& & Point-wise &
145 & 10.9 & $< 0.1$ & 4.4 & 0.9 & & 44.9 &
157 & 25.6 & $< 0.1$ & 6.3 & 3.1 & & 27.3 \\
{Optimizer} & & & 
1219 & 4.0 & $< 0.1$ & 1.1 & 0.3 & & 30.6 &
1219 & 3.9 & $< 0.1$ & 1.1 & 0.5 & & 31.3 \\
{Copies / Transposes} & & & 
535 & 104.9 & - & 63.2 & 8.6 & & 66.9 &
708 & 213.2 & - & 92.6 & 26.1 & & 48.3 \\
{Allreduce (NCCL)} & & & 
35 & 56.4 & $< 0.1$ & 0.6 & 4.6 & & 1.2 &
30 & 58.7 & $< 0.1$ & 0.6 & 7.2 & & 1.1 \\
{Type Conversions} & & &
&&&&&&&
201 & 1.3 & - & 0.6 & 0.2 & & 51.3 \\
GPU Idle & &&
&&&&&&&
    & 14.2 &  &     & 1.7 & & \\
%Idle Time & & & & ??.? & - & - \\
\hline
Total     & & &
3170 & 1215.9 & 14.41 & 220.9 & & 75.5 & 20.2 &
3497 & 817.3 & 28.82 & 203.6 & & 28.2 & 27.7
\\ \hline
\end{tabular}%
}

\caption{\label{fig:single-node-perf-deeplab}%
Detailed single node performance analysis of DeepLabv3+ network training for FP32 (left) and FP16 (right) precision.  Kernels are grouped by category, with the total time, FLOPs, and memory traffic reported for each.  The fraction of time spent in kernels from each category is shown along with the fraction of peak math and memory performance achieved by kernels in that category. Values reported are subject to some measurement uncertainty (see text).}
\end{figure}

\end{document}